\documentstyle[12pt]{article}
\begin{document}
\title{A possible glueball contribution to the Goldberger-Treiman relations.
\thanks{Work supported in part by the KBN-Grant 2-0224-91-01.}}
\author{Jan Bartelski\\
Institute of Theoretical Physics, Warsaw Uniwersity,\\
Ho$\dot{z}$a 69, 00-681 Warsaw, Poland. \\ \\
\and
Stanislaw  Tatur \\
N.Copernicus Astronomical Center,\\ Polish Academy of Sciences,\\
Bartycka 18, 00-716 Warsaw, Poland.}
\date{}
\maketitle
\begin{abstract}
\noindent We discuss the influence of glueball coupling to nucleons on the weak
axial-vector coupling constants including singlet channel. We consider
a possibility of introduction of constituent gluon contribution to the proton
spin. The estimated value for this quantity seems to be rather small.
\end{abstract}
The EMC experiment \cite{ash} started a great interest in the problem of
the proton spin. Analyzing naively, the quark contribution to this quantity
 came out unexpectadely small (see e. g. ref.\cite{BEK}). In the one of the
interpretations authors assumed \cite{alt} that there exists additional
gluon contribution which nearly cancel the quark one, making the measured
result rather small.
In this context the role of the axial anomaly and the Goldberger-Treiman
relations was also discussed \cite{ven,efr,jbst}. These relations are
very important because they allow to calculate the axial-vector coupling
constants which are directly related to the quark contribution to the nucleon
spin. Especially crutial, because of the existence of the axial-vector
anomaly, is the Goldberger-Treiman relation for the singlet axial-vector
current.

In this paper, introducing phenomenologically the mixing of a glueball state
with pseudoscalar mesons $\eta$ and $\eta'$, we want to discuss a possible
glueball contribution to the Goldberger-Treiman relation for the singlet
axial-vector current. Naively one would expect that the glueball coupling to
nucleons measures in some sense a gluonic content of nucleon spin, in analogy
to
the quark case. We introduce a constituent gluon contribution to the
nucleon spin and we try to estimate its value.
We also speculate about a glueball contribution to the gluonic content of a
proton spin and we compare the result with the perturbative gluon contribution
needed to understand the EMC experimental result.

Taking into account SU(3) mass breakings, together with isospin mass breaking
and $\pi$-$\eta$-$\eta'$ mixing, one can obtain Goldberger-Treiman
relations for the third, eights and singlet (in the generalized version)
component of the axial-vector current \cite{efr,jbst}. We will write the
relations for these components in the vector form. Introducing properly
normalized vector of weak axial-vector coupling constants, pseudoscalar meson
coupling constants to nucleons and the quantity $\tilde{G}$ (it sums the pole
contributions from the physical particles $\pi$, $\eta$ and $\eta'$) we have:
\begin{equation}
\tilde{g}_{A}=\frac{f_{\pi}}{2M} M^{2}_{sym} \tilde{G} ,
\end{equation}
where $f_{\pi}$ is a pion decay constant ($f_{\pi}\simeq 132\; MeV$),
$M_{sym}$-mass matrix in the SU(3) basis, whereas:
\begin{equation}
\tilde{g}_{A}=
\left(
\begin{array}{c}
\frac{1}{\sqrt{2}}g^{(3)}_{A}\\
\frac{1}{\sqrt{6}}g^{(8)}_{A}\\
\frac{1}{\sqrt{3}}\Delta\Sigma
\end{array}
\right),\hspace{5mm}
\tilde{g}_{NN}=
\left(
\begin{array}{c}
g_{\pi NN}\\
g_{\eta NN}\\
g_{\eta'NN}
\end{array}
\right),\hspace{5mm}
\tilde{G}=
\left(
\begin{array}{c}
G^{(3)}\\
G^{(8)}\\
G^{(0)}
\end{array}
\right) .
\end{equation}
We have also for $x$=3,8,0:
\begin{equation}
G^{(x)}=\sum_{p=\pi,\eta,\eta'}\frac{\Omega_{px}}{m^{2}_{p}}g_{pNN} ,
\end{equation}
or writing eq.(3) in more compact form:
\begin{equation}
\tilde{G}=\hat{\Omega}^{\dagger}(M^{2}_{phys})^{-1}\tilde{g}_{NN}  ,
\end{equation}
where $M_{phys}$ is a diagonal mass matrix with physical masses and orthogonal
matrix $\hat{\Omega}$ connects physical and SU(3)(third, eighth and singlet)
states. Hence, the eq.(1) can be rewritten in the form:
\begin{equation}
\tilde{g}_{A}=\frac{f_{\pi}}{2M}M^{2}_{sym}\hat{\Omega}^{\dagger}
(M^{2}_{phys})^{-1} \tilde{g}_{NN}=\frac{f_{\pi}}{2M}\hat{\Omega}^{\dagger}
\tilde{g}_{NN},
\end{equation}
where the second equality follows because the matrix $\hat{\Omega}$
transforms also mass matrices i.e.: $\hat{\Omega} M^{2}_{sym}\hat{\Omega}
^{\dagger}=M^{2}_{phys}$.The final result (eq.(5)) does not involve meson
masses,
only a mixing angles and could be considered as a generalization,  for the
diagonal octet states, of an old Goldberger-Treiman relation for pion. The
mixing of $\pi$ with $\eta$ and $\eta'$, due to the isospin breaking, is
negligible when compared with SU(3) breaking and we will neglect it.
However, we would like to extend above relations by taking into account gluon
degrees of freedom and include glueball mixing with $\eta$ and $\eta'$.

For a long time $\eta(1440)$, called also $\iota$, was considered as a glueball
candidate. As was pointed out \cite{m3} there are some problems with this
interpretation. Recent Mark III results has changed the experimental situation
showing that $0^{-+}$ state at 1440 MeV is not a single resonance but rather
a mixture of three different states. We will consider one of these states,
namely $0^{-+}$ state with the mass around 1490 MeV, as a possible glueball
candidate and call it as before $\iota$. Our main reasoning depends on
existence of a glueball and not on its particular mass.
One of the possible demonstration of pseudoscalar glueball existence would
be a mixing with pseudoscalar $q\bar{q}$ states. We will use for this mixing
a model discussed by us previously \cite{jbt}. Old models of $\eta$ ,
$\eta'$ mixing with glueball (see e.g. ref.\cite{jbt}) do not take into
account a new experimental situation. Let us assume that the physical states
$\eta$, $\eta'$ and $\iota$ can be expressed in terms of the SU(3) states
$\eta_{8}$, $\eta_{0}$ and a pure psedoscalar glueball G using the orthogonal
matrix $\hat{\Omega}$ i.e.
\begin{equation}
\left(
\begin{array}{c}
\eta\\
\eta'\\
\iota
\end{array}
\right)=
\hat{\Omega}
\left(
\begin{array}{c}
\eta_{8}\\
\eta_{0}\\
G
\end{array}
\right) .
\end{equation}
As was pointed before this matrix diagonalizes also the (mass)$^{2}$  matrix
for pseudoscalar SU(3) states. Using information about the mass matrix from
the quark model (with inclusion of chiral corrections for $m^{2}_{88}$),
assuming $m^{2}_{G8}=0$ and taking into account experimental information from
$\iota\!\rightarrow\! 2\gamma$ decays which give $\mid\Omega_{\eta'G}\mid^{2}
\cong0.075$ we can calculate the mixing matrix $\hat{\Omega}$  \cite{jbt}:
\begin{equation}
\hat{\Omega}=
\left(
\begin {array}{rrr}
0.94&0.34&\pm 0.07\\
-0.34&0.90&\pm 0.27\\
\pm0.03&\mp 0.28&0.96
\end{array}
\right) .
\end{equation}
There is an arbitrariness in the sign of some matrix elements and we choose
upper sign in order to get a proper sign of gluon spin contribution. We have
shown in ref.\cite{jbt} that such model is in agreement with all available
informations about the radiative decays of pseudoscalar mesons.
We expect from the interpretation of the EMC effect that not only quarks but
also gluons play an important role in the spin structure of a nucleon.
Considering the mixing of pseudoscalar states built out of quarks with a
glueball built out of gluons means  that we take into account additional
(independent of u, d and s) gluonic degrees of freedom. We assume that in
addition to eighth component and singlet quark currents there exists a gluonic
current which divergence, in analogy to the quark ones, is given by a linear
combination of considered fields with coefficients determined by the masses
in the third row of the $M^{2}_{sym}$ matrix (see eq.(1)). It is not clear
to us how such current should be constructed in terms of fundamental fields.
We assume the mixing of very different objects, the states that in the chiral
limit are massless Goldstone bosons (therefore the one particle approximation
in Goldberger-Treiman relation is justified) and gluon-antigluon bound states.
We will use for the gluonic current  the assumption of the domination by the
glueball state (we are concious that it is not well justified)
and neglect the higher as well as multiparticle states. We hope that our
estimate gives if not whole than at least a part of a gluonic contribution.
The one particle contribution, corresponding to the glueball, is proportional
to $f_{G}$ which need not be the same as $f_{\pi}$. There exists however a
model
where $f_{G}=f_{\pi}$ \cite{ros} and in order to estimate glueball contribution
we will consider this case at the begining. Introducing in our case:
\begin{equation}
\tilde{g}_{A}=
\left(
\begin{array}{c}
\frac{1}{\sqrt{6}}g^{(8)}_{A}\\
\frac{1}{\sqrt{3}}\Delta\Sigma\\
g^{G}_{A}
\end{array}
\right),\hspace{5mm}
\tilde{g}_{NN}=
\left(
\begin{array}{c}
g_{\eta NN}\\
g_{\eta'NN}\\
g_{\iota NN}
\end{array}
\right) ,
\end{equation}
in a very similar way as in the case of SU(3) quark axial-vector currents we
will get the following relations:
\begin{eqnarray}
g^{(8)}_{A}&=&\sqrt{6}\frac{f_{\pi}}{2M}(\Omega_{\eta8}g_{\eta
NN}+\Omega_{\eta'8}
g_{\eta'NN}+\Omega_{\iota8}g_{\iota NN}) ,  \\
\Delta\Sigma&=&\sqrt{3}\frac{f_{\pi}}{2M}(\Omega_{\eta0}g_{\eta
NN}+\Omega_{\eta'0}
g_{\eta' NN}+\Omega_{\iota0}g_{\iota NN}) ,  \\
g^{(G)}_{A}&=&\hspace{5.5mm}\frac{f_{\pi}}{2M}(\Omega_{\eta G}g_{\eta
NN}+\Omega_{\eta'G}
g_{\eta' NN}+\Omega_{\iota G}g_{\iota NN}) .
\end{eqnarray}
Taking $g_{\eta NN}=6.8$, $g_{\eta' NN}=7.3$ from \cite{dumb} and matrix
elements of our mixing matrix $\hat{\Omega}$ we get numerically:
\begin{eqnarray}
g^{(8)}_{A}&=&\Delta u+\Delta d-2\Delta s=0.67 \pm 0.003g_{\iota NN} , \\
\Delta\Sigma&=&\Delta u+\Delta d+\Delta s=1.08 \mp 0.03g_{\iota NN} , \\
g^{G}_{A}&=&\Delta g=\pm 0.17+0.07g_{\iota NN} ,
\end{eqnarray}
where the signs exibit an arbitrariness in our solution (see eq.(7)).
The quantities $g^{(8)}_{A}$ and $\Delta\Sigma$ being combinations of
$\Delta u$, $\Delta d$ and $\Delta s$ describe the constituent quark content
of nucleon spin and in analogy we can consider $g^{G}_{A}$
as a constituent gloun contribution to the nucleon spin, namely $\Delta g$.
The value of $g_{\iota NN}$ is not known from the experiment. Because of this
we present our results in the Table 1 showing the dependence of considered
quantities on $g_{\iota NN}$.
\begin{center}
Table 1\\ [12pt]
\end{center}
The dependence of the weak axial-vector coupling constants: $g^{(8)}_{A}$,
$\Delta\Sigma$ and $\Delta g$ on values of glueball-nucleon coupling constant
$g_{\iota NN}$.
\begin{center}
\begin{tabular}{|c||c|c|c|}\hline
$g_{\iota NN}$&3&5&7\\
\hline\hline
$g^{(8)}_{A}$&0.68&0.69&0.71\\
$\Delta\Sigma$&0.98&0.91&0.84\\
$\Delta g$&0.38&0.51&0.64\\
\hline
\end{tabular}
\end{center}
{}From equations (12-14) and
the Table 1 we see that the contribution from glueball does not influence the
values of $g^{(8)}_{A}$ and $\Delta\Sigma$ very much (there were atempts
\cite{chin} to explain the EMC data taking big $g_{\iota NN}$). For example
for not so small value of $\iota$-nucleon coupling $g_{\iota NN}=5$ we have
$g^{(8)}_{A}=0.69$ and $\Delta\Sigma=0.91$. The obtained values are not very
different from the values obtained previously \cite{jbst} for $\theta_{p}=
-20^{o}$ without mixing with the glueball state and $g^{(8)}_{A}$ is close
to the value gotten from experimental figures: $0.58\pm0.03$ (using
$(g_{A}/g_{V})_{N\rightarrow P}$ and $(g_{A}/g_{V})_{\Sigma^{-}\rightarrow N}$
from \cite{pdg}) or $0.60\pm 0.12$ (estimate given by Jaffe and Manohar
\cite{jaffe}). The obtained value of $\Delta g$ is rather small and even for
relatively large $g_{\iota NN}=5$  we get only $\Delta g=0.51$. This value
is of course for low energy scale, say  $\mu^{2}\approx 0.3\; GeV^{2}$.
Using type of reasoning proposed by authors of ref.\cite{reya}
we will try to estimate what should be the value of $\Delta g$ at low energy
scale in order to understand results of EMC experiment at $Q^{2}_{EMC}=10.7\;
GeV^{2}$. We define $\Delta \tilde{g}=N_{F}\frac{\alpha_{s}}{2\pi}\Delta g$
($N_{F}=3$) and evolution equations approximately give:
\begin{equation}
\Delta\tilde{g} (\mu^{2})\cong \Delta\tilde{g} (Q^{2}_{EMC})
\end{equation}
Calculating $\Delta\tilde{g}(Q^{2}_{EMC})$ from the "experimental" value for
$G_{1}(0)=0.13\pm 0.17$ (see e.g. ref.\cite{jaffe}) and $\Delta\Sigma=0.91$
we get:
\begin{equation}
\Delta\tilde{g} (Q^{2}_{EMC})=-G_{1}(0)+\Delta\Sigma=0.85\pm 0.17
\end{equation}
Using $\alpha_{s}(Q^{2}_{EMC})\cong 0.25$
we obtain from eq.(16) $\Delta g(Q^{2}_{EMC})=6.5$, and hence, taking that
$\alpha_{s}(\mu^{2})/\alpha_{s}(Q^{2}_{EMC})=3.55$ \cite{reya}, we have:
\begin{equation}
\Delta g(\mu^{2})\cong\frac{\alpha _{s}(Q^{2}_{EMC})}{\alpha
_{s}(\mu^{2})}\Delta
 g(Q^{2}_{EMC})=1.8
\end{equation}
This means that what we have got is (for $g_{\iota NN}=5$) about four
times smaller then the value needed to explain the EMC experiment. The value
of the constituent gluonic spin contribution as measured by the interaction
with glueball is not very big. In other words using our $\Delta\Sigma=0.91$
and $\Delta g=0.51$ we get $G_{1}(0)\cong0.77$. To avoid the conflict with the
experiment we need to introduce a large perturbative gluonic contribution
$\Delta g$ so that $\Delta\tilde{\Sigma}=\Delta\Sigma-N_{F}\frac{\alpha _{s}}
{2\pi}\Delta g$ could be equal to the EMC value $0.13\pm 0.17$.
In principle we can take for $g_{\iota NN}$ value as high as 13.6 (for example
the condition $\Delta s=0$, i.e. $g^{(8)}_{A}=\Delta\Sigma$ for $\mu^{2}=0.3\;
GeV^{2}$ gives $g_{\iota NN}=10.4$) and explain all the needed glueball
contribution but we consider such large value as unreasonable. Let us make
one more comment. Because we have in the divergences of the axial-vector
currents the mixing of very different objects i.e.: glueball and nearly
massless Goldstone bosons there is no a priori reason that in
Goldberger-Treiman relations $f_{G}$ is equal to $f_{\pi}$. In the case of
$f_{G}\neq
f_{\pi}$ we modify our formulae replacing $G^{(G)}$ by
$(f_{G}/f_{\pi})G^{(G)}$.
Hence, we get:
\begin{eqnarray}
g^{(8)}_{A}&=&g^{(8)}_{A}(f_{G}=f_{\pi})\\
\Delta\Sigma&=&\Delta\Sigma(f_{G}=f_{\pi})+\sqrt{3}\frac{f_{\pi}}{2M}(\frac{f_{G}}
{f_\pi}-1)m^{2}_{G0}G^{(G)}\\
\Delta g&=&\Delta g(f_{G}=f_{\pi})+\frac{f_{\pi}}{2M}(\frac{f_{G}}
{f_\pi}-1)m^{2}_{GG}G^{(G)}
\end{eqnarray}
Now, the results for $\Delta\Sigma$ and $\Delta g$ depend much stronger than
before on $g_{\iota NN}$ coupling.
Let us make a speculation and take as an example $f_{G}/f_{\pi}=1.5$. We
obtain:
\begin{eqnarray}
\Delta\Sigma&=&1.00-0.04g_{\iota NN}\\
\Delta g&=&0.45+0.10g_{\iota NN}
\end{eqnarray}
For $g_{\iota NN}=5$ we get $\Delta\Sigma=0.78$ and $\Delta g=0.95$, and the
last figure should be compared with the value of $\Delta g=1.48$ obtained
from the evolution equations (eq.(17)). In this case the value of $\Delta g$
is only about two third of the needed value and is still too small.

We have shown that the hypothetical glueball coupling with nucleons does not
influence very much the values of $g^{(8)}_{A}$ and $\Delta\Sigma$.
With a rather speculative assumptions we have tried to estimate the constituent
gluon spin contribution as measured by the glueball interaction with the
nucleons.
The obtained figures are only a small part of the ones needed to understand
the value of proton spin as measured in the EMC experiment.


\newpage
\begin{thebibliography}{99}
\bibitem{ash} J. Ashman et al.,Phys. Lett. {\bf B206} (1988) 364; Nucl. Phys.
{\bf B328} (1989) 1.
\bibitem{BEK} S. Brodsky, J. Ellis, M. Karliner, Phys. Lett. {\bf B206}
(1988) 309.
\bibitem{alt} G. Altarelli, G. Ross, Phys. Lett. {\bf B212} (1988) 391; \\
A. V. Efremov, O. V. Teryaev, Dubna raport JINR, EZ-88-297 (1988), unpublished.
\bibitem{ven} G. Veneziano, Mod. Phys. Lett. {\bf A4} (1989) 1605; \\
G. H. Shore, G. Veneziano, Phys. Lett. {\bf B244} (1990) 75;\\
T. Hatsuda, Nucl. Phys. {\bf B329} (1990) 376;\\
A. V. Efremov, J. Soffer, N. A. T\" ornqvist, Phys. Rev.
Lett. {\bf 64} (1990) 1495,
\bibitem{efr}A. V. Efremov, J. Soffer, N. A. T\" ornqvist, Phys. Rev. {\bf D44}
(1991) 1369.
\bibitem{jbst} J.Bartelski, S.Tatur, Phys.Lett. {\bf B 265} (1991) 192.
\bibitem{m3} Mark III Collaboration, Z.Bai et al., Phys.Rev.Lett. {\bf 65}
(1990) 2507.
\bibitem{jbt} J. Bartelski, S. Tatur, Phys. Lett. {\bf B289} (1992) 429,
\bibitem{ros} C.Rosenzweig, A.Salomone, J.Schechter, Phys.Rev.
{\bf D 24} (1981) 2545.
\bibitem{dumb} O. Dumbrajs et al., Nucl. Phys. {\bf B216} (1983) 277.
\bibitem{chin} K. Chao, J. Wen, H. Zeng, CERN preprint CERN-TH. 6288/91 (1991),
unpublished.
\bibitem{pdg} Particle Data Group, Phys. Rev. {\bf D45} (1992) vol.11.
\bibitem{jaffe} R. L. Jaffe, A. Manohar, Nucl. Phys. {\bf B337} (1990) 509.
\bibitem{reya} M. Gl\" uck, E. Reya, Phys. Lett. {\bf B270} (1991) 65.
\end{thebibliography}
\end{document}